\documentclass[jkps,preprint,twocolumn,fleqn,showpacs,showkeys,10pt]{revtex4-1}
\usepackage{graphicx}
\usepackage{amssymb}
\usepackage{amsmath}
\usepackage{bm}
\begin{document}
\setcounter{page}{0}
\title[]{Critical configurations for a system of semidegenerate fermions}
\author{Carlos R. \surname{Arg\"{u}elles}}
\email{carlos.arguelles@icranet.org}
\author{Remo \surname{Ruffini}}
\affiliation{ICRANet, Piazzale della Repubblica 10-65122, Pescara, Italy\\
ICRA, Department of Physics, Sapienza University of Rome, Italy}
\author{Bernardo M.O. \surname{Fraga}}
\affiliation{ICRA, Department of Physics, Sapienza University of Rome, P.le Aldo Moro, 5 I-00185, Rome, Italy\\
Universit\'e de Nice Sophia Antipolis, Grand Chateau Parc Valrose Nice CEDEX 2, France}

\date[]{}

\begin{abstract}
We study an isothermal system of semi-degenerate self-gravitating fermions in general relativity. Such systems present mass density solutions with a central degenerate core, a plateau and a tail which follows a power law behaviour $r^{-2}$. The different solutions are governed by the free parameters of the model: the degeneracy and temperature parameters at the center, and the particle mass $m$. We then analyze in detail the free parameter space for a fixed $m$ in the keV regime, by studying the one-parameter sequences of equilibrium configurations up to the critical point, which is represented by the maximum in a central density ($\rho_0$) Vs. core mass ($M_c$) diagram. We show that for fully degenerate cores, the known expression for the critical core mass $M_c^{cr}\propto m_{pl}^3/m^2$ is obtained, while instead for low degenerate cores, the critical core mass increases showing the temperature effects in a non linear way. The main result of this work is that when applying this theory to model the distribution of dark matter in galaxies from the very center up to the outer halos, we do not find any critical core-halo configuration of self-gravitating fermions, which be able to explain the super massive dark object in their centers together with an outer halo simultaneously.
\end{abstract}



\maketitle

\section{INTRODUCTION}

Systems of self-gravitating semi-degenerate fermions in general relativity were studied in \cite{rr} and more recently with applications to dark matter in galaxies in \cite{charly}. It was shown that, for a given central temperature parameter ($\beta_0$) in agreement with the corresponding observed halo circular velocity, there are lower bounds for the central degeneracy parameter ($\theta_0$) and particle mass ($m\gtrsim0.4$ keV) above which the \textit{observed} halo mass and radius are fulfilled. The density profiles solutions in this approach present a novel core-halo morphology composed by a quantum degenerate core followed by a low degenerate plateau until they reach the $r^{-2}$ Boltzmannian regime. This interesting overall morphology provides the flat rotation curves in the outermost part of the galaxies as well as a possible alternative to massive black holes in their centers (see \cite{rrchina} and \cite{firstwork}).

The system of Einstein equations are written in a spherically symmetric space-time metric $g_{\mu \nu}={\rm diag}(e^{\nu},-e^{\lambda},-r^2,-r^2\sin^2\theta)$,
where $\nu$ and $\lambda$ depend only on the radial coordinate $r$, together with the thermodynamic equilibrium conditions of Tolman \cite{tolman}, and Klein \cite{klein},
\begin{equation}
e^{\nu/2} T=const.\, , \quad e^{\nu/2}(\mu+m c^2)=const, \nonumber
\end{equation}
where $T$ is the temperature, $\mu$ the chemical potential, $m$ the particle mass and $c$ the speed of light. We then write the system of Einstein equations in the following dimensionless way,
\begin{align}
		&\frac{d\hat M}{d\hat r}=4\pi\hat r^2\hat\rho \label{eq:1}\\
		&\frac{d\theta}{d\hat r}=\frac{\beta_0(\theta-\theta_0)-1}{\beta_0}
    \frac{\hat M+4\pi\hat P\hat r^3}{\hat r^2(1-2\hat M/\hat r)}\\
    &\frac{d\nu}{d\hat r}=\frac{\hat M+4\pi\hat P\hat r^3}{\hat r^2(1-2\hat M/\hat r)} \\
    &\beta_0=\beta(r) e^{\frac{\nu(r)-\nu_0}{2}}\, . \label{eq:2}
\end{align}
The variables of the system are the mass $M$, the metric factor $\nu$ , the temperature parameter $\beta=k T/(m c^2)$ and the degeneracy parameter $\theta=\mu/(k T)$. The dimensionless quantities are: $\hat r=r/\chi$, $\hat M=G M/(c^2\chi)$, $\hat\rho=G \chi^2\rho/c^2$ and $\hat P=G \chi^2 P/c^4$, with $\chi=2\pi^{3/2}(\hbar/mc)(m_p/m)$ and $m_p=\sqrt{\hbar c/G}$ the Planck mass.  The mass density $\rho$ and pressure $P$ are given by Fermi-Dirac statistics (see also \cite{charly}).

This system is solved for a fixed particle mass $m$ in the keV range, with initial conditions $M(0)=\nu(0)=0$, and given parameters $\theta_0>0$ (depending on the chosen central degeneracy), and $\beta_0$. We thus construct a sequence of different thermodynamic equilibrium configurations where each point in the sequence has different central temperatures $T_0$ and central chemical potential $\mu_0$, so that satisfy the $\theta_0$ fixed condition.

\begin{figure}[!hbtp]
\centering
\includegraphics[width=\linewidth]{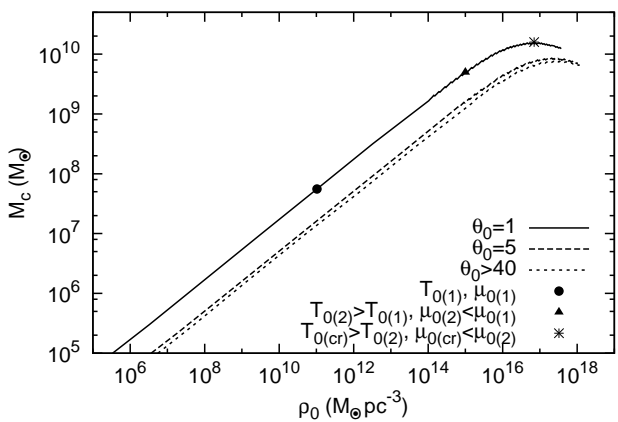}
\caption{Different sequences of equilibrium configurations plotted in a central density ($\rho_0$) Vs. core mass ($M_c$) diagram. The critical core mass is reached at the maximal value of $M_c$. Each sequence is built for selected values of $\theta_0=\mu_0/kT_0$ and different values of $T_0, \mu_0$ varying accordingly.}
\label{fig:1}
\end{figure}

Defining the core radius $r_c$ of each equilibrium system at the first maximum of its rotation curve, or equivalently at the degeneracy transition point in which $\theta(r_c)=0$, we represent the results obtained for each sequence in a central density ($\rho_0$) vs. core mass ($M_c$) diagram (see Fig.\ref{fig:1}). It is shown that the critical core mass $M_c^{cr}$ is reached at the maximum of each $M_c(\rho_0)$ curve.

It is important to emphasize that we are not interested in follow the history of equilibrium states of one specific system. Thus, the standard stability analysis as done for compact stars or in dense stellar cluster (see e.g. \cite{rrcluster}), which is based on the constancy of the entropy per nucleon (S/N) along the equilibrium sequence of a given configuration, does not apply here. Nonetheless, in computing the $M_c(\rho_0)$ curves in Fig.~\ref{fig:1} we have explored the full range of $\theta_0>1$ and $\beta_0>10^{-10}$ parameters (including the critical ones). Then the equilibrium sequences with constant specific entropy (S/N), which differ from the ones with constant $\theta_0$ considered here, necessarily must be contained within the full ($T_0,\mu_0$) parameter space covered in Fig.~\ref{fig:1}.

In Table~\ref{table:1} we show a set of central critical parameters of the model together with the correspondent critical core masses, for a very wide range of fixed central degeneracy parameters $\theta_0$ and $m=8.5$ keV$/c^2$.

\begin{table}
\begin{ruledtabular}
\begin{tabular}{cccc}
$\theta_0$ & $\beta_0^{cr}$ & $\mu_0^{cr}/mc^2$ & $M_c^{cr} (M_\odot)$ \\
\colrule
 1 & $6.45\times10^{-2}$ & $6.45\times10^{-2}$ & $1.59\times10^{10}$ \\
 5 & $2.23\times10^{-2}$ & $1.11\times10^{-1}$ & $7.91\times10^9$ \\
 40 & $8.33\times10^{-3}$ & $3.33\times10^{-1}$ & $7.44\times10^9$ \\
 55 & $6.06\times10^{-3}$ & $3.33\times10^{-1}$ & $7.44\times10^9$ \\
 100 & $3.33\times10^{-3}$ & $3.33\times10^{-1}$ & $7.44\times10^9$ \\
\end{tabular}
\caption{Critical temperature parameter and normalized chemical potential at the center of each different critical configuration, for different fixed central degeneracies.}
\end{ruledtabular}
\label{table:1}
\end{table}

Defining the halo radius of each configuration at the onset of the flat rotation curve, we show in Table II the critical halo magnitudes $r_h^{cr}$, $M_h^{cr}$ and $v_h^{cr}$ corresponding to the same set of critical parameters as given in Table I.

\begin{table}
\begin{ruledtabular}
\begin{tabular}{cccc}
$\theta_0$ & $r_h^{cr} (pc)$ & $M_h^{cr}/mc^2 (M_\odot)$ & $v_h^{cr} (km/s)$ \\
\colrule
 1 & $4.4\times10^{-1}$ & $5.7\times10^{11}$ & $7.5\times10^4$ \\
 5 & $4.0\times10^{-1}$ & $4.3\times10^{11}$ & $6.2\times10^4$ \\
 40 & $4.3\times10^{3}$ & $1.1\times10^{15}$ & $3.3\times10^4$ \\
 55 & $2.9\times10^{5}$ & $6.0\times10^{16}$ & $2.9\times10^4$ \\
 100 & $2.0\times10^{11}$ & $2.3\times10^{22}$ & $2.2\times10^4$ \\
\end{tabular}
\caption{Critical halo magnitudes of different critical configurations, for different fixed central degeneracies as given in Table I.}
\end{ruledtabular}
\label{table:2}
\end{table}

The results obtained in Tables I and II imply a marked division in two different families depending on the value of $M_c^{cr}$.

\textit{i}) The first family: the critical mass has roughly a constant value $M_c^{cr}=7.44\times10^9 M_\odot$. This family corresponds to large values of the central degeneracy ($\theta_0\geq40$), where the critical temperature parameter is always lower than $\beta_0^{cr}\lesssim8\times10^{-3}$ and the critical chemical potential $\mu_0^{cr}\approx$ const. Physically, these highly degenerate cores are entirely supported against gravitational collapse by the degeneracy pressure. In this case the critical core mass is uniquely determined by the particle mass according the relation $M_c^{cr}\propto m_{pl}^3/m^2$ (see also section III).

\textit{ii}) The second family: the critical core mass increases from $M_c^{cr}=7.44\times10^9 M_\odot$ up to $M_c^{cr}\sim10^{10} M_\odot$. This case corresponds to critical cores with a lower central degeneracy compared with the former family ($1<\theta_0<40$). Here the critical temperature parameter ($\beta_0\sim10^{-2}$), is closer to the relativistic regime with respect to the first family. This result physically indicates that the thermal pressure term has now an appreciable contribution to the total pressure, which supports the critical core against gravitational collapse. In this case $M_c^{cr}$ is completely determined by the particle mass $m$, the central temperature $T_0^{cr}$ and the central chemical potential $\mu_0^{cr}$ (see section III).

\begin{figure}
 \centering
\includegraphics[width=.9\linewidth]{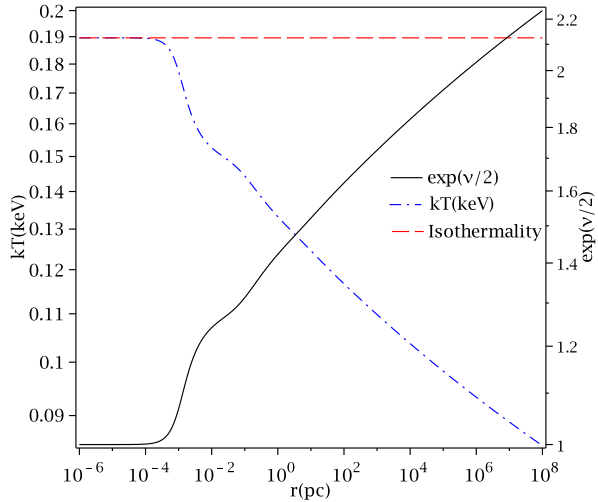}
\caption{The critical temperature profile of the system (in keV) and the critical metric, for $\theta_0=5$ and $\beta_0^{cr}=2.23\times 10^{-2}$. The dashed line corresponds to the isothermality condition, $Te^{\nu/2}=const$.}
\label{fig:2}
\end{figure}

\begin{figure}
 \centering
\includegraphics[width=.9\linewidth]{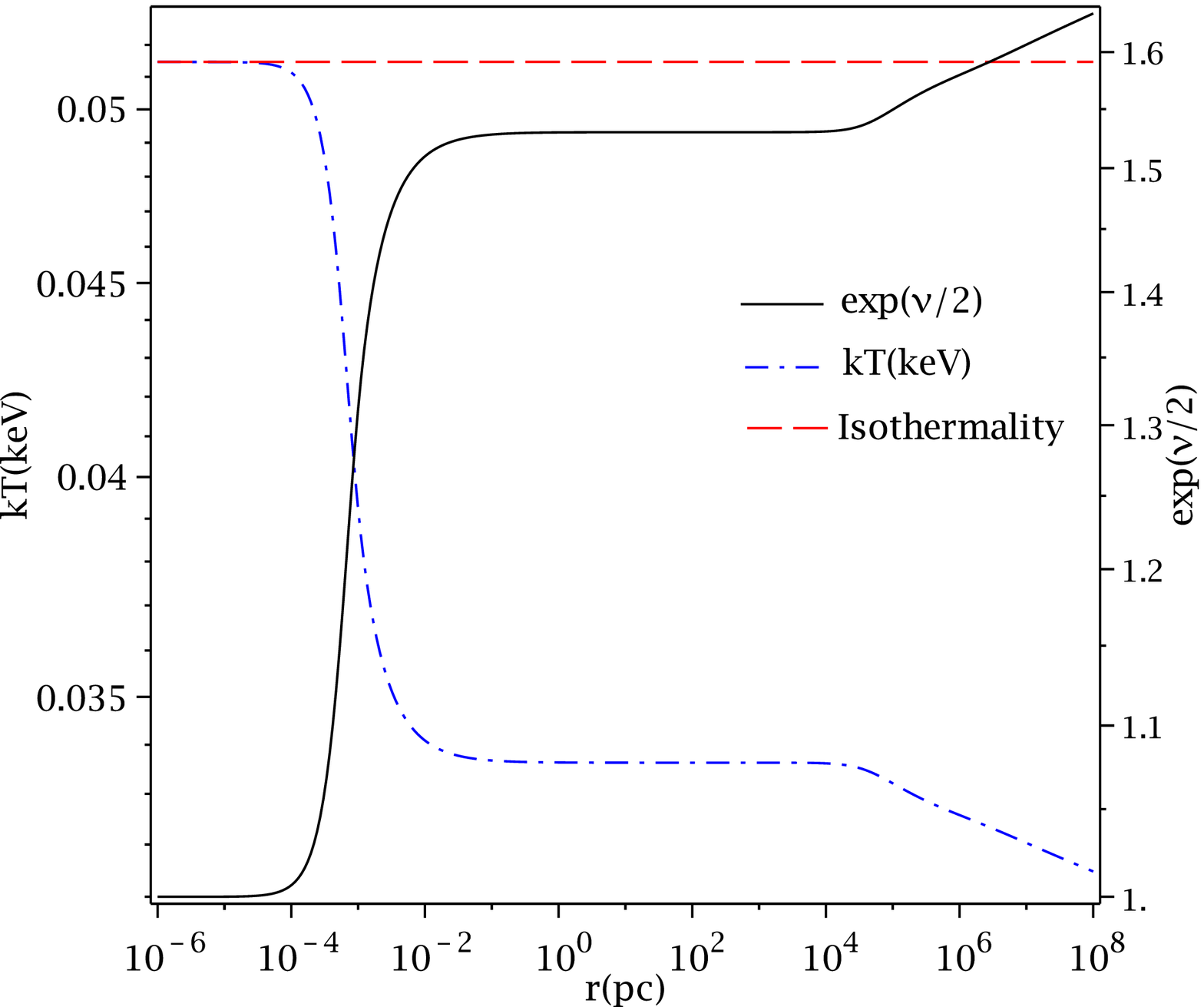}
\caption{The critical temperature of the system (in keV) and the critical metric, for $\theta_0=55$ and $\beta_0^{cr}=6.06\times10^{-3}$. The red line corresponds to the isothermality condition, $Te^{\nu/2}=const$.}
\label{fig:3}
\end{figure}

In Figs.~(\ref{fig:2}) and (\ref{fig:3}) we show a critical metric factor $e^{\nu/2}$ and a critical temperature $kT$ as function of the radius for the two different families mentioned above.

\section{Astrophysical application}
We will now attempt to use the critical configurations obtained before to explain the DM distribution in galactic halos, as well as providing an alternative candidate to the standard central black hole paradigm. From now on, we will use a fixed particle mass of $m=$8.5 keV$/c^2$, being this choice motivated by the fact we want to deal with super massive dark objects having critical core masses of the order $M_c^{cr}\propto m_{pl}^3/m^2\sim10^9 M_{\odot}$.

\begin{figure}[!hbtp]
\centering
\includegraphics[width=1.\linewidth]{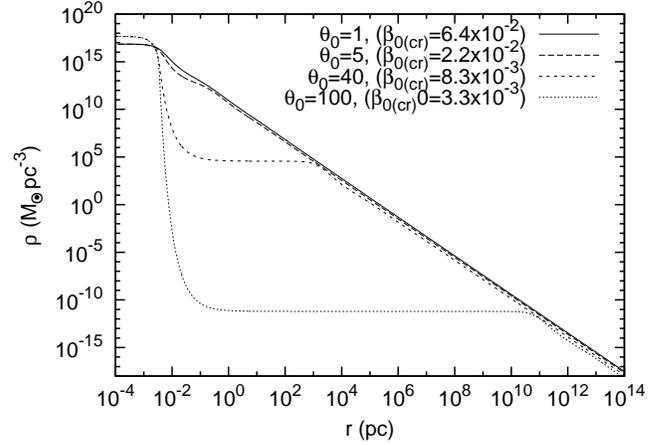}
\caption{Critical density profiles for different values of $\theta_0$ with the correspondent critical temperature parameters $\beta_0^{cr}$.}
\label{fig:4}
\end{figure}

\begin{figure}[!h]
\centering
\includegraphics[width=1.\linewidth]{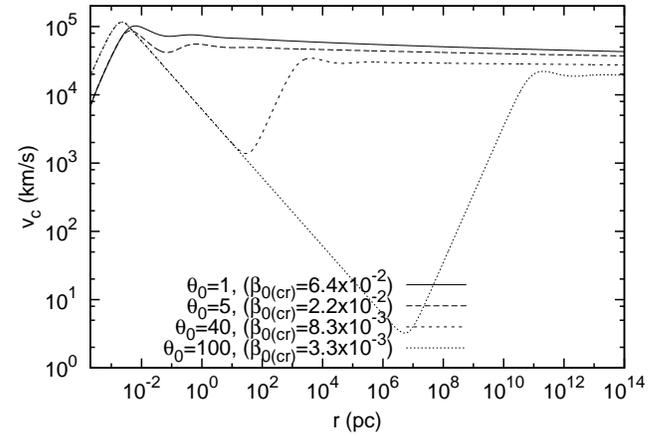}
\caption{Critical rotation curves for different values of $\theta_0$ as given in Fig.~\ref{fig:4}. To note the high values of $v_c(r)\sim10^4$ km/s in the flat parts of each curve due to the high critical temperature parameters $\beta_0^{cr}$.}
\label{fig:5}
\end{figure}

\begin{figure}[!h]
\centering
\includegraphics[width=1.\linewidth]{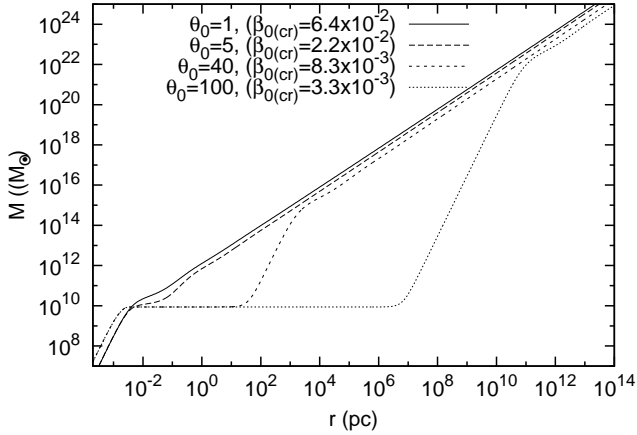}
\caption{Critical mass profiles for different values of $\theta_0$ as given in Figs.~\ref{fig:4}--\ref{fig:5}.}
\label{fig:6}
\end{figure}

In Figs.~(\ref{fig:4}), (\ref{fig:5}) and (\ref{fig:6}) we show different critical density profiles, critical rotation curves and critical mass profiles respectively, for a wide range of different central degeneracy parameters.

From Fig.~\ref{fig:5} and Table II we see that critical configuration of self-gravitating fermions with central degeneracies ranging from $\theta_0=1$ (i.e. mainly thermal pressure supported cores) up to $\theta_0=100$ (i.e. mainly degeneracy pressure supported cores), have flat rotation velocities $v_h^{cr}\sim10^4$ km/s. This is, well above any observed value. However, since larger values of $\theta_0$ imply lower values for $\beta_0^{cr}$, as shown in Table I, there is a point (at $\theta_0\sim 10^6$ and so $\beta_0^{cr}\sim10^{-7}$) where the \textit{halo} rotation curves reach the typical observed value of $v_h^{cr}\sim10^2$ km/s. Nonetheless, since even for $\theta_0=100$ we see (Fig.~\ref{fig:5}) that the \textit{halo} radius is larger than 10 Gpc with total critical mass larger than $10^{20}\,M_{\odot}$, this is again well above than any observed galactic halo. We then conclude that any critical configuration belonging to this model is not able to reproduce the halo of galaxies.

Moving now to a completely different size scale, this is $r<10^{-2}$ pc, even though at this scales the critical core masses are of the order of the more massive supermassive black holes in the center of Active Galactic Nuclei (AGN) (see Table I), the outer part of the system is either too large or have extremely large velocities compared to observations as explained before. Even inferred maser velocities on AGNs, with values up to 1000 km/s on a sub-parsec scale are always below the critical velocity for any value of $\theta_0$ as seen from Fig.~\ref{fig:5}. We conclude therefore that critical configurations of semi-degenerate self-gravitating fermions cannot be used to model AGNs.


 \section{An analytical expression for the critical mass}
It is useful to find an analytical formula for the critical mass to try to understand the physics behind it. For this we will
use the Newtonian hydrostatic equilibrium equation corresponding to the last stable configuration, where the pressure due to gravity is balanced by a high relativistic semi-degenerate Fermi gas :
{\small
\begin{align}
P_g(r)&=&P_T^{ur}(r), \nonumber \\
\frac{GM(r)\rho(r)}{r}&\approx&\frac{\mu^4}{12\pi^2(\hbar c)^3}+\frac{\mu^2(kT)^2}{6\sqrt\pi(\hbar c)^3},
\label{lastequil}
\end{align}
}
where $P_T^{ur}(r)$ is the ultra relativistic approximation of a highly relativistic Fermi gas ($\mu\gg mc^2$), which has been expanded up to second order in temperature around $\mu/kT\gg1$ (see e.g. \cite{landau}). We have used in \ref{lastequil} the fermi energy ($\epsilon_f=\mu$) with the rest energy substracted-off in consistency with the theoretical formulation of our model.
Considering that the density in the cores is nearly constant (see Fig.~\ref{fig:4}), i.e., $\rho=\rho_0^{cr}\approx const., \forall r\leq r_c^{cr}$, we can write the core radius as $r_c^{cr}=(3M_c^{cr}/(4\pi\rho_0^{cr}))^{1/3}$. With this, we can rewrite (\ref{lastequil}) as follows,
{\small
\begin{align}
\left(\frac{4\pi}{3}\right)^{1/3}G(M_c^{cr})^{2/3}(\rho_0^{cr})^{4/3}\approx \notag \\
\approx\frac{\mu^4}{12\pi^2(\hbar c)^3}\left(1+\frac{2\pi^2}{\theta_0^2}\right),
\label{lastequil1}
\end{align}
}
Finally, we write the central mass density as $\rho_0^{cr}\approx n^{ur}\,m$, where $n^{ur}=\mu^3/(3\pi^2(\hbar c)^3)$ is the ultra-relativistic particle number density. With this expression for $\rho_0^{cr}$ in (\ref{lastequil1}) we can directly give $M_c^{cr}$ in terms of $\theta_0$ as:
\begin{equation}
M_c^{cr}\approx\frac{3\sqrt{\pi}}{16}\frac{M_{pl}^3}{m^2}\left(1+\frac{2\pi^2}{\theta_0^2}\right)^{3/2}.
\label{Mcranalyt}
\end{equation}

It is clear from this equation that for high central degenerate systems ($\theta_0\gg\sqrt{2}\pi$), the critical core mass $M_c^{cr}$ is independent of $\theta_0$ and then proportional to $M_{pl}^3/m^2$. However, for low values of the central degeneracy ($\theta_0\sim\sqrt{2}\pi$) the second term in (\ref{Mcranalyt}) starts to be relevant, showing the finite temperature effects. In fact, using $\theta_0=40$, we have $M_c^{cr}=7.62\times 10^9\,M_{\odot}$, just a 2\% difference with the numerical result of $7.44\times 10^9\,M_{\odot}$. However, for $\theta_0=5$ we have $M_c^{cr}=1.79\times 10^{10}\,M_{\odot}$, almost a factor 2 above the numerical value of $7.91\times 10^9\,M_{\odot}$ obtained in Table I. This shows that our approximation of an ultra-relativistic fermi gas in newtonian equilibrium breaks down and a fully relativistic treatment is needed.

\begin{acknowledgments}
BF is supported by the Erasmus Mundus Joint Doctorate Program by grant number 2010-1816 from the EACEA of the european union. We acknowledge the hospitality at APCTP where part of this work was done.
\end{acknowledgments}

\end{document}